\documentclass[noeprint,twocolumn,amsmath,amssymb,floatfix,superscriptaddress,pre,prl]{revtex4-2}

\usepackage{bm}
\usepackage{bbm}
\usepackage{graphicx, amsmath, xcolor}
\usepackage{physics}
\usepackage{dcolumn}
\usepackage{comment}
\usepackage{siunitx}
\usepackage{amsmath,amssymb,amsfonts}
\usepackage[colorlinks=true,linkcolor=red, citecolor=blue]{hyperref}
\usepackage{booktabs}

\definecolor{cerisepink}{rgb}{0.93, 0.23, 0.51}

\begin{document}

\title{Boundaries Program Deformation in Isolated Active Networks}

\author{Zixiang Lin}
\affiliation{Global College, Shanghai Jiao Tong University, Shanghai, People’s Republic of China}
\author{Shichen Liu}
\affiliation{Division of Biology and Biological Engineering, California Institute of Technology, Pasadena, CA, USA}
\author{Shahriar Shadkhoo}
\affiliation{Division of Biology and Biological Engineering, California Institute of Technology, Pasadena, CA, USA}
\author{Jialong Jiang}
\affiliation{Division of Biology and Biological Engineering, California Institute of Technology, Pasadena, CA, USA}
\author{Heun Jin Lee}
\affiliation{Department of Applied Physics, California Institute of Technology, Pasadena, CA, USA}
\author{David Larios}
\affiliation{Division of Biology and Biological Engineering, California Institute of Technology, Pasadena, CA, USA}
\author{Chunhe Li}
\affiliation{Global College, Shanghai Jiao Tong University, Shanghai, People’s Republic of China}
\author{Hongyi Bian}
\affiliation{Global College, Shanghai Jiao Tong University, Shanghai, People’s Republic of China}
\author{Anqi Li}
\affiliation{Global College, Shanghai Jiao Tong University, Shanghai, People’s Republic of China}
\author{Rob Phillips}
\affiliation{Division of Biology and Biological Engineering, California Institute of Technology, Pasadena, CA, USA}
\affiliation{Department of Applied Physics, California Institute of Technology, Pasadena, CA, USA}
\affiliation{Department of Physics, California Institute of Technology, Pasadena, CA, USA}
\author{Matt Thomson}
\affiliation{Division of Biology and Biological Engineering, California Institute of Technology, Pasadena, CA, USA}
\author{Zijie Qu}
\email{zijie.qu@sjtu.edu.cn}
\affiliation{Global College, Shanghai Jiao Tong University, Shanghai, People’s Republic of China}

\date{\today}

\begin{abstract}
Cellular structures must organize themselves within strict physical constraints, operating with finite resources and well-defined boundaries. Classical systems demonstrate only passive responses to boundaries, from surface energy minimization in soap films to strain distributions in elastic networks. Active matter fundamentally alters this paradigm - internally generated stresses create a bidirectional coupling between boundary geometry and mass conservation that enables dynamic control over network organization. Here we demonstrate boundary geometry actively directs network deformation in reconstituted microtubule-kinesin systems, revealing a programmable regime of shape transformation through controlled boundary manipulation. A coarse-grained theoretical framework reveals how boundary geometry couples to internal stress fields via mass conservation, producing distinct dynamical modes that enable engineered deformations. The emergence of shape-preserving and shape-changing regimes, predicted by theory and confirmed through experiments, establishes boundary geometry as a fundamental control parameter for active materials. The control principle based on boundaries advances both the understanding of biological organization and enables design of synthetic active matter devices with programmable deformation.
\end{abstract}

\maketitle

\section{Introduction}

Boundaries fundamentally dictate material behavior across physical systems, but their role transforms dramatically when moving from passive to active matter \cite{marchetti2013hydrodynamics, needleman2017active, xu2023geometrical, watwani2025influence}. In classical systems, boundaries act as passive constraints—soap films minimize surface area \cite{taylor1976structure, almgren1976geometry}, elastic networks distribute strain uniformly \cite{landau2012theory}, and fluid flows simply conform to container geometry. In contrast, active matter systems generate internal stresses that dynamically interact with their confining interfaces. This bidirectional coupling means that boundaries can actively direct the organization of the material, rather than just limit it \cite{marchetti2013hydrodynamics, needleman2017active}. Mechanical properties of active networks have been extensively studied, experimentally \cite{foster2015active, mizuno2007nonequilibrium, thoresen2011reconstitution, kohler2011structure, kohler2012contraction, schuppler2016boundaries, foster2017connecting, suzuki2017spatial} and theoretically \cite{kruse2005generic, liverpool2009mechanical, prost2015active, lee2001macroscopic, liverpool2003instabilities, kruse2004asters, aranson2005pattern, liverpool2006rheology, juelicher2007active, mackintosh2008nonequilibrium, koenderink2009active, gao2015multiscale, ronceray2016fiber, gladrow2016broken, furthauer2019self}, but boundary geometry is rarely used as a primary parameter. Understanding how boundary geometry influences active matter behavior is essential for predicting and controlling material deformation in both synthetic and biological contexts.

Biological systems serve as the major source of inspiration in designing synthetic active systems \cite{peraza2014origami, pinson2017self, furthauer2021design}. In cells, cross-linked polymer networks mediate the active forces that are generated by motor proteins through hydrolyzing ATP. \textit{In vitro} experiments with cell extracts or reconstituted networks of Microtubules (MTs), and kinesin motor proteins show self-organization into structures including asters and contractile/extensile networks \cite{weisenberg1984atp, verde1991taxol, ndlec1997self, surrey2001physical, sanchez2012spontaneous}. The organization behavior of active cytoskeletal networks is involved in mitosis and meiosis, cell dynamics, morphogenesis, and signal transduction \cite{goldstein1999road, rice1999structural, sharp2000microtubule, hirokawa2009kinesin, li2012novel, shelley2016dynamics}. Furthermore, during certain phases of the cell cycle, such as mitosis, the concentration of MT polymers has been observed to remain relatively stable \cite{loiodice2019quantifying}. It suggests the conservation of mass of the MT-kinesin active networks in the contractile process.

Early experiments have demonstrated an optically controlled active matter system composed of MTs and engineered kinesin motor proteins \cite{ross2019controlling}. However, recent studies have revealed two major limitations in achieving complete control of active networks: The background solution containing floating MTs will flow into the active network, making it difficult to maintain a stable mass \cite{ross2019controlling, qu2021persistent}, and the active network is accompanied by persistent fluid flow fields, so that interactions between the network and its environment remain largely uncontrolled \cite{qu2021persistent, lemma2023spatio, yang2025dynamic}. These two limitations also make it difficult to define the boundaries of the active network, which prevents us from further exploring the role of the boundaries in the self-organization of the active network. In addition, the boundary has emerged as a critical factor in shaping the topology of active matter systems \cite{duclos2020topological, shankar2022topological, boguna2021network, bowick2022symmetry}.

\begin{figure*}[t]
        \includegraphics[width=\linewidth]{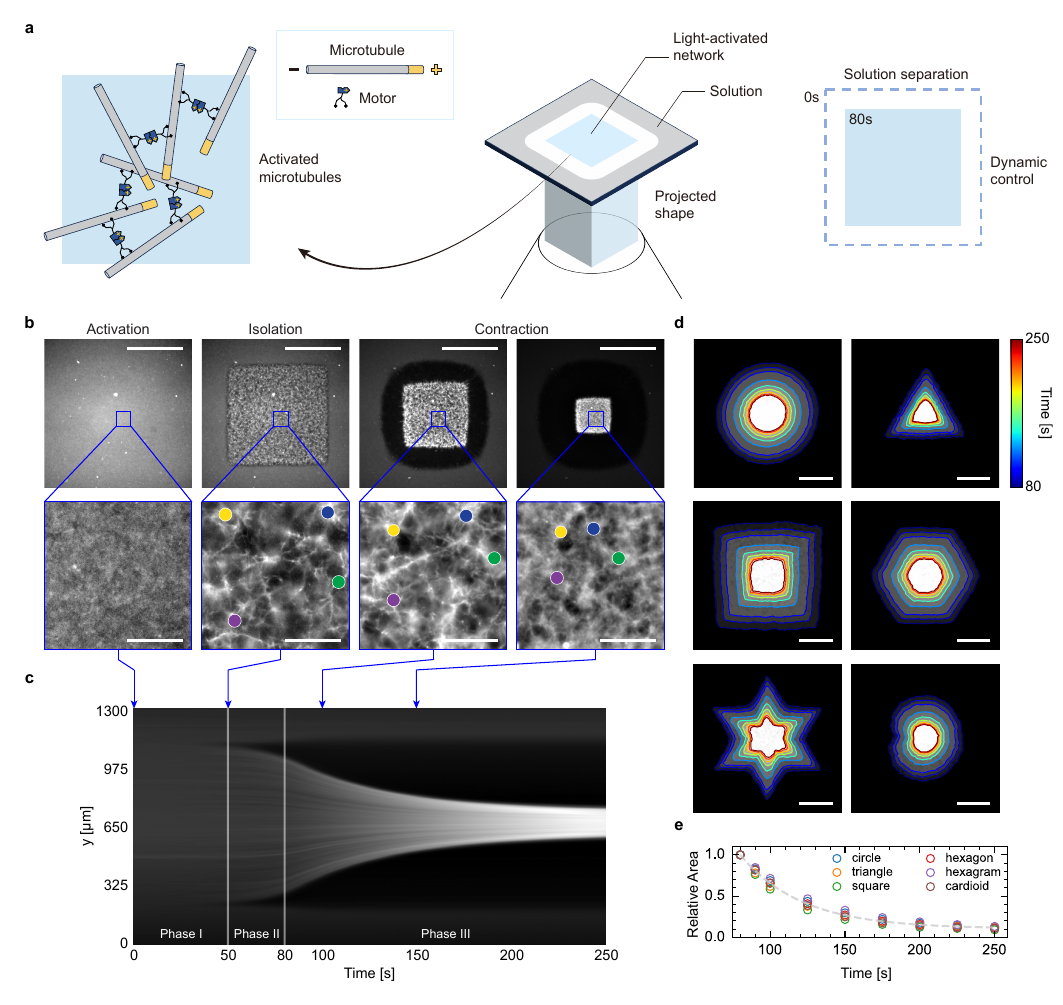}
        \caption{
         Optical-control protocol first activates cross-linking motor proteins to form the MT networks and isolates the network from embedding solution, allowing them to contract. (a) shows the active matter system used in this study, which consists of fluorescently labeled, stabilized MT filaments and kinesin motors that cross-link under illumination. An initial pulse of light activates motor proteins within a region of illumination. Activated motor proteins crosslink the MTs and form a contractile network. Isolation of the network from the solutions requires a second pulse at around $\sim 50-80$s. (b) shows the macroscopic (top row) and microscopic (second row) snapshots of the network, from left to right: during the activation and network formation, at the time of isolation, and shape preserving contraction. The colored dots in the second row track the loci of four distinct microscopic asters in time. Scale bar for top row, $500 \mu$m; Scale bar for second row, $50 \mu$m. (c) shows the profile of the contracting network in time (horizontal axis). The major three phases of the dynamics are separated by white vertical lines. (d) For six different boundary geometries the contraction of networks is portrayed by overlaying the networks’ boundaries as they shrink in time. Scale bar, $500 \mu$m. (e) The relative area of networks in (d) decays over time. The gray dashed curve is an exponential fit to the data, with $\bar{\chi} = 0.89 \pm 0.01, \bar{\tau} = (41 \pm 2)$s.
         } 
        \label{fig:fig-1}
\end{figure*}

In this work, we develop an optical control protocol to activate motor proteins within a region of illumination, form active MT-motor networks, and isolate them from the surrounding solution. Our strategy utilizes a recently developed optical experimental system to form and isolate active networks of different geometries \cite{ross2019controlling}. The dynamics of isolated networks are dominated by active stresses with fluid drag \cite{guntas2015engineering,ross2019controlling} (Fig. \ref{fig:fig-1}a). Across several distinct geometries, we observed that active networks undergo boundary contraction deformations. We introduce a theoretical dynamics of an isolated active MT-motor network and show that the deformation is a direct consequence of viscous-like active stresses and passive boundary forces. Our model provides insights into programming active stresses through controlled modulation of light patterns and intensities. Specifically, we design protocols for spatiotemporal modulations of light intensity to achieve static bending as well as temporally alternating bending directions in the network. By demonstrating that engineered boundary geometry can serve as a universal control parameter to program complex deformation of MT-motor networks, our work advances the fundamental understanding of the active matter system and paves the way for designing synthetic materials with programmable deformation.

\section{Activity Preserves the Boundary Geometry Memory of Contracting Networks}

We first studied the whole contractile process using a square, $900{\mu}{\text{m}}$ illumination (Fig. \ref{fig:fig-1}b, top row, Supplementary Video 1). A combination of microscopy and image analysis was used to track and infer network dynamics using labeled MTs (Fig. \ref{fig:fig-1}b, second row). Our observations show that the contracting network exhibits completely different behaviors at different stages throughout the contraction process (Supplementary Information); therefore, we have divided it into three phases (Fig. \ref{fig:fig-1}c). In phase I ($t \lesssim 50\si{s}$), the contractile MT-motor network forms, and its shape is determined by the region of illumination. The activated network is isolated from the background solution by the end of this phase. In phase II ($50\si{s} \lesssim  t \lesssim 80\si{s}$), the isolated network starts to contract with a increasing speed. The area of the network decreases over time while the density of the cross-linked network increases. At the end of this phase, the network's contraction speed reaches the maximum value of the whole contraction process. In phase III ($t\gtrsim 80\si{s}$), the network persistently contracts with a decreasing speed. This phase is the main process of network contraction and lasts the longest. The network shrinks from an area close to the initial illuminated area to an area only about $10$\% of its original size. Our subsequent discussion will mainly focus on this phase. Deceleration of contraction as the density of filaments increases, and thus the MT-MT steric interactions increase. Eventually, the network approaches a contraction limit. During the contractile phases (II and III), the network approximately retains the initial geometry of the light pattern both in macroscope and microscope (Fig. \ref{fig:fig-1}b).

Performing experiments on several distinct geometries reveals universal dynamics that shed light on the underlying active mechanism. We next studied contracting circles as well as polygonal networks (triangles, squares, and hexagons) of different sizes; $450,\,600,\,750$ and $900{\mu}{\text{m}}$. In comparison to convex geometries that are identified by uniformly positive boundary curvature, concave geometries have arcs of both positive and negative curvature, which provides a more stringent test for verifying the activity-dominated contraction. Therefore, we also prepared networks in two concave geometries: hexagrams and cardioids. All the networks with different geometries show an approximately self-similar contractile behavior (Fig. \ref{fig:fig-1}d, Supplementary Video 2-7). In order to quantitatively assess self-similarity, we first segment images to find the regions occupied by the networks at different times (Supplementary Information). Next, for the network at two time points $t_1$ and $t_2 > t_1$, with areas $A_1$ and $A_2 < A_1$, we scale down the larger area by $\sqrt{A_2/A_1}$, and align the centers of the two geometries. Self-similarity is defined as the ratio of the bitwise overlap area $A_1 \& A_2$, and $A_2$. To account for stochastic rigid rotations of each network around its center of mass, we maximize the self-similarity with respect to relative rotations over the range of $(-20,+20)$ degrees. The deviation from self-similarity, $\delta(t_1, t_2)$, is calculated by subtracting the self-similarity from unity. Across all networks examined we found that $\delta \in (2\%, 10\%)$ over the entire course of the dynamics. High degree of persisting boundary geometry preservation suggests spatially-uniform and isotropic contraction of the networks. It also phenomenally suggests a linear radial velocity field in the network.

To further test whether boundary geometry has an effect on the contractile rate, we tracked the area of networks during the phase III of the contraction (Fig. \ref{fig:fig-1}e). In our experiments, the initial MT density is kept constant across networks of different sizes and shapes. The relative area, defined as the ratio of the area over time $A(t)$ to the initial area $A_0$, decays exponentially from $1$ across different geometries. i.e., $A(t)/A_0 = \chi e^{-t/\tau} + 1 - \chi$, where $\chi, \tau$ are constants. It is noticed that the time scale $\tau$ is approximately the same for different networks which is inversely proportional to the activity (Fig. \ref{fig:fig-1}e). Therefore, the boundary geometry only defines the domain of the activity but is irrelative to its strength. Given that activity is an increasing function of the light intensity, we expect the contraction to speed up upon cranking up the intensity.

\section{Theoretical Model Reveals Mechanism of Universality of Contraction}

Programming active contractile networks requires quantitative understanding of the response of the system to the external probes, e.g. light in our experiments. To understand how contractions emerge in response to internally generated stress, we developed and analyzed a coarse-grained theoretical model of active networks. Our phenomenology draws on the following experimentally grounded postulates: (1) Isotropicity within the network: the initially randomly oriented MTs organize small asters that are connected to each other via some intermediate MTs. The asters are, however, connected in random directions. Therefore, for length scales of multiple asters, size isotropicity seems to be a reasonable assumption (Fig. \ref{fig:fig-1}b, second row). (2) Activated motor proteins induce contractile stress. (3) Steric interactions become progressively stronger as the network contracts and balance out the contractile stress at an equilibrium density of the network.

The dynamics of the system is governed by the conservation laws of total mass and momentum, where total refers to the MT network and the fluid. Mass conservation demands $\partial_t(\rho_n + \rho_f) = -\div{(\rho_n\vb*{v}_n + \rho_f \vb*{v}_f )} = 0$, where $\rho_{n/f}$ are network/fluid densities. We dropped the network’s subscript hereafter. Neglecting the inertial terms on macroscopic time scales, momentum conservation (force balance) for the network requires $\vb*{f}_d + \vb*{f}_v + \vb*{f}_e = \div{\vb*{\sigma}}$. Here the passive external force exerted from the surrounding fluid on the network is $\vb*{f}_d = -\gamma(\vb*{v} - \vb*{v}_f)$, in which $\gamma$ is the effective drag coefficient. The viscous response of the network to the total stress is $\vb*{f}_v = \eta\laplacian{\vb*{v}}$, in which $\eta$ is the effective network viscosity. For the elastic response of the network, due to the long enough relevant time scale $\tau$, the stress is considered as $\vb*{f}_e \approx 0$ \cite{foster2015active}. Under the assumption of $|\vb*{v}_f|\ll|\vb*{v}|$, the governing equations are
\begin{subequations}{\label{eq:main}}
\begin{align}
&\partial_t\,\rho + \div(\rho\vb*{v}) = 0,    \\
&\eta\laplacian{\vb*{v}} - \gamma\vb*{v} = \div{\vb*{\sigma}}.
\end{align}
\end{subequations}

The expression for the active contractile velocity field $\vb*{v}_a$ is obtained by considering both the continuity equation and the fraction contracted of the network. We found that the emission intensity of MTs is approximately uniformly distributed within the active network for the entire contraction process (Supplementary Information). Under the assumption that the local density of the network $\rho(\vb*{r})$ is proportional to the intensity of emission light captured in gray-scale images $\mathcal{I}(\vb*{r})$, the density of the active network is always nearly uniform, i.e., $\rho(\vb*{r}, t) \approx \rho(t)H(\mathcal{R}(\arg{(\vb*{r})}) - |\vb*{r}|)$, where the Heaviside step function is defined as $H(x)=\mathbf {1} _{x\geq 0}$, the boundary is expressed by $\mathcal{R}(\theta)$, and $\vb*{r} = (r, \theta) = (|\vb*{r}|, \arg(\vb*{r}))$ in the polar coordinates. The simplified equation implies a linear radial flow field and the exponential relaxation fraction of the network area corresponds to the magnitude decaying over time. Taken together, the macroscopic contractile behavior is interpreted by the active flow
\begin{equation}{\label{eq:va}}
    \vb*{v}_a(\vb*{r}, t) =  -\frac{\vb*{r}}{2\tau}\frac{\chi e^{-(t-T_c)/\tau}}{\chi e^{-(t-T_c)/\tau} + 1-\chi}H(\mathcal{R}(\arg{(\vb*{r})}) - |\vb*{r}|),
\end{equation}
where $T_c$ is the lag time before phase III begins (Supplementary Information).

The dependency of the active stress on the intensity of light is crucial to programming the dynamics of the network. In order to understand this dependency, we next solve the momentum equation within the active contractile velocity field $\vb*{v}_a$. The internal active stress is assumed to be isotropic, namely proportional to the identity matrix $\mathbbm{1}$. In 2D we consider $\vb*{\sigma}^a = \frac{1}{2}\tr(\vb*{\sigma}^a)\mathbbm{1} = \sigma^a\mathbbm{1}$. Then, the stress is solved as
\begin{equation}{\label{eq:stress}}
    \vb*{\sigma}^a = s\rho(\rho - \rho_\infty)\mathbbm{1},
\end{equation}
where $s$ is the strength of active stress, and $\rho_\infty$ represents the final density of the network in phase IV (Supplementary Information). The relation between stress and MT density $\vb*{\sigma}^a\propto \rho(\rho - \rho_\infty)$ can be interpreted as two effects \cite{verde1991taxol, elting2014force, surrey2001physical, foster2015active, ndlec1997self}: (1) the collection of MTs minus the kinesin motor mediated ends to drive a contractile stress that is proportional to the density; (2) the extensile stress driven by steric interactions which is quadratic in the density.

\begin{figure*}[t]
        \includegraphics[width=\linewidth]{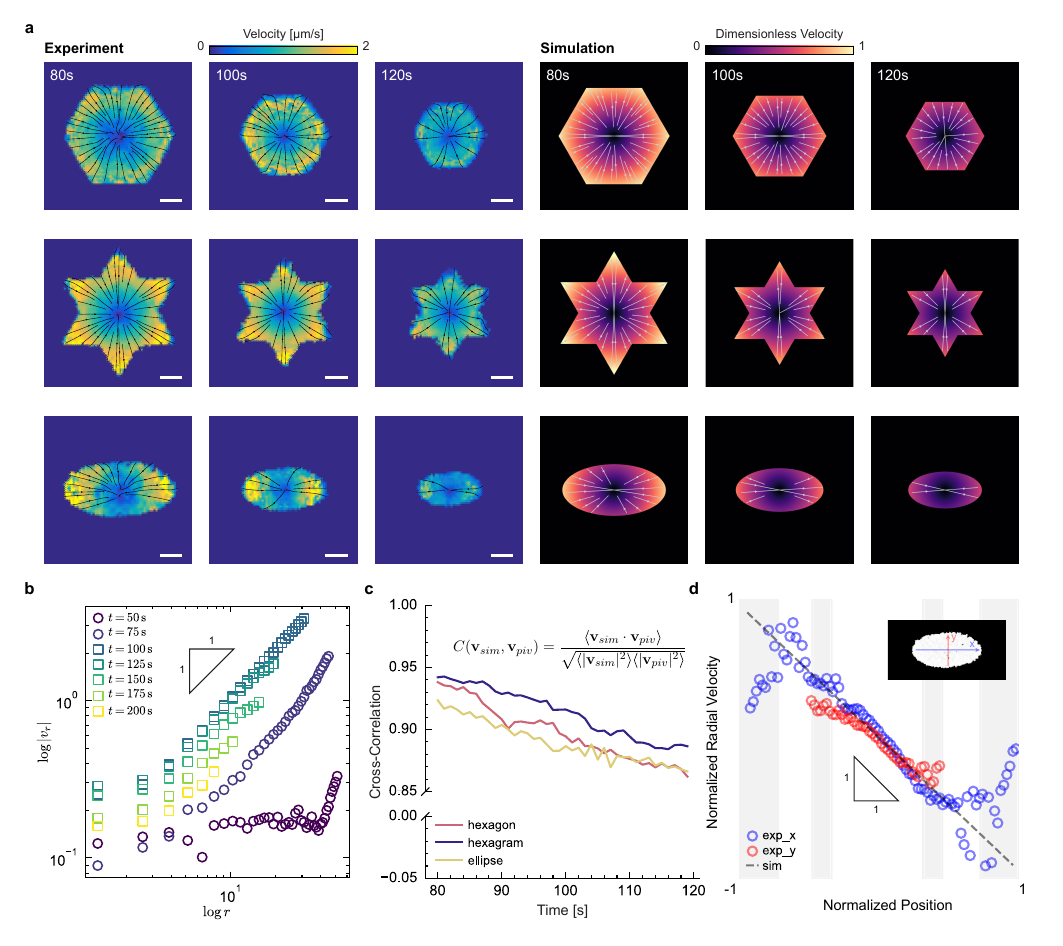}
        \caption{
         Comparison and agreement between experiments and theory supports the role of activity-induced viscous interaction in mass-conserved contraction. (a) For three geometries of hexagon, hexagram, and ellipse, the flow fields depicted with streamlines are shown as extracted via PIV in experiments (left) and simulated (right). The velocity fields are approximately linear and radial in all cases. Scale bar, 100$\mu$m. (b) shows the relation between the radial velocity $v_r$ with the distance from the mass center $r$. The velocity fields are non-linear in phase II (circles), but become strongly linear in phase III (squares). (c) shows the cross-correlations between the simulated velocity fields $\vb*{v}_{sim}$, and the experimental velocity fields by PIV analyses $\vb*{v}_{piv}$ over the course of the network’s contraction. The cross-correlation, defined in (c), where $\langle f \rangle = \int\dd^2x\,f(\vb*{x})$, is normalized, hence bounded between $[-1, +1]$. The correlations decrease due to the decreasing network areas and fixed precision of PIV analyses (Supplementary Information). (d) The radial velocity distribution of an ellipse network at $100\si{s}$ in (a). The radial velocity is normalized as $\tilde{v}_r = (2v_r - \max{v_r} - \min{v_r}) / (\max{v_r} - \min{v_r})$, and the position is normalized as $\tilde{r}_x = r_x / a, \tilde{r}_y = r_y / b$, where $a, b$ are the lengths of semi-major axis and semi-minor axis, hence both the normalized radial velocity and position are bounded between $[-1, +1]$. The black dashed line shows the simulation result, which follows a power-law scaling of $-1$. The shaded regions (outer along the major axis and inner along the minor axis) delineate the near-boundary zones where the experimental data deviate from the simulation due to the boundary mechanics.
         } 
        \label{fig:fig-2}
\end{figure*}

The results of the simulations reproduce the same dynamics as observed in experiments. The velocity field extracted by Particle Image Velocimetry (PIV) from contracting networks \cite{raffel2018particle}, and those obtained from simulations are both linear and radial over geometries and over time (Fig. \ref{fig:fig-2}a-b). To validate our simulation results, we calculate the correlation coefficients of the simulated velocity fields and the experimental data inferred by PIV, which have a lower bound $\mu_C \simeq 0.85$ but decrease over time (Fig. \ref{fig:fig-2}c). To further investigate the reasons why the correlation decreases over time, we extracted the velocity field on the two axes of an ellipse network. We found that the velocity is strongly linear with the position in the internal region of the network but shows a trend of gradually increasing from the outside to the inside in the regions near the boundary (Fig. \ref{fig:fig-2}d). Combined with the existence of the deviation from self-similarity $\delta$, we suppose that there is another velocity field $\vb*{v}_p$ that causes the boundary geometry to distort.

\begin{figure*}[t]
        \includegraphics[width=\linewidth]{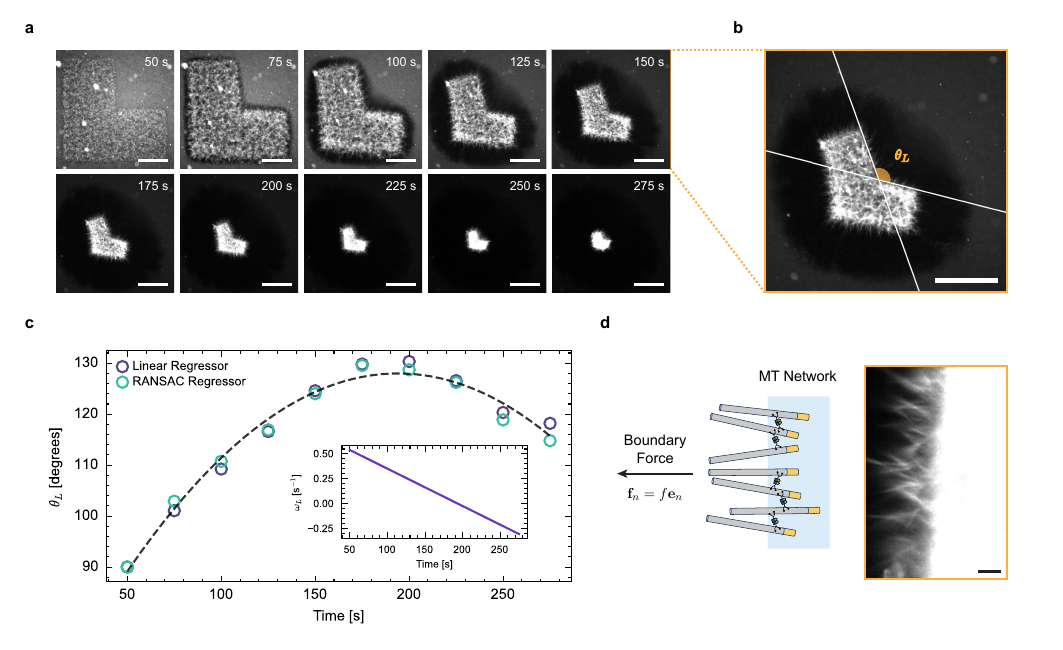}
        \caption{
         Boundary mechanics provides forces, guiding the deformation of the active network. (a) shows the contractile behavior of a MT network in L-shape. Scale bar, 200$\mu$m. (b) shows the measurement of the tracked angle $\theta_L$. Two white straight lines represent the fit lines of two edges by the regression methods. Scale bar, 200$\mu$m. (c) shows the tracked angle $\theta_L$ changes with time. Linear regression (green) and random sampling and consensus (RANSAC) regression (purple) are used to measure the angle respectively (Supplementary Information). The data is well fitted by a quadratic function $\theta_L = (-0.0019 \pm 0.0001)t^2 + (0.73 \pm 0.03)t + (57.5 \pm 2.2)$ in degrees (gray dashed line). The corresponding angular velocity $\omega_L = (-0.0038 \pm 0.0002)t + (0.73 \pm 0.03)$ (inset). (d) shows the schematic and experimental snapshot of MT network boundary. The outermost MTs are arranged with a strong perpendicular orientation to the boundary, performing the passive response of the network. Scale bar, 10$\mu$m.
         } 
        \label{fig:fig-3}
\end{figure*}

To expand the degree of distortion of the boundary geometry to obtain the velocity field $\vb*{v}_p$, we prepared L-shape illuminations and tracked the changes in the right angle of networks over time (Fig. \ref{fig:fig-3}a-b, Supplementary Video 8). The tracked angle $\theta_L$ changes significantly during the phase III, with the maximum derivation $\max(\delta\theta_L/\theta_{L0}) \approx 45\%$ and a linearly decreasing angular velocity $\omega_L$ (Fig. \ref{fig:fig-3}c). Observing the experimental network contraction process, it is noticed that the behavior of the outermost MTs is different from that of the inner MTs. The orientation of the outermost MTs is almost completely perpendicular to the boundary, rather than the disordered appearance of the inner MTs (Fig. \ref{fig:fig-3}d). The possible reason is that the light-activated motors on the outermost MTs are partly activated or non-activated and they can only interact with the inner networks. These MTs perform a resistance force to the inner MT network, which is perpendicular to the boundary, i.e., $\vb*{f}_{n} = f\vb*{e}_n$, where $\vb*{e}_n$ represents the unit normal vector on the boundary $\mathcal{R}$. For closed boundaries $\sum_\mathcal{R} \vb*{f}_{n} = 0$, but the local torque is non-zero for non-symmetric shapes like L-shape, i.e., $M_{local} = \sum_{local} \vb*{r}\times \vb*{f}_{n} \neq 0$. It implies an inhomogeneous angular velocity field $\vb*{\omega}$. At low Reynolds numbers, the velocity field is obtained by the 2D Helmholtz decomposition \cite{chwang1975hydromechanics},
\begin{equation}{\label{eq:vp}}
    \vb*{v}_p = \alpha\Delta(\mathcal{R}(\arg{(\vb*{r})}) - |\vb*{r}|)\vb*{e}_n + \vb*{\omega}\times\vb*{r} + \beta\vb*{r},
\end{equation}
where $\alpha$ represents the magnitude of flows caused by the perpendicular forces, the delta function $\Delta$ is defined as $\Delta(x) = \mathbf {1} _{x = 0}$, and $\beta$ is a constant defined with $\div{\vb*{v}_p} = 0$ to satisfy the conservation of mass (Supplementary Information).

The total velocity field is the superposition of the active self-affine contraction velocity field and the passive divergence-free deformation velocity field,
\begin{align}{\label{eq:v-final}}
    {\vb*{v}} = {\vb*{v}_a} + {\vb*{v}_p} =  
     &(-\frac{\tilde{\chi}\tilde{H}}{2\tau}r + \beta r + \frac{{\tilde{\alpha}}{\mathcal{R}}} {\sqrt{\mathcal{R}^2 + {\mathcal{R}'}^2}})\vb*{e}_r \notag \\
     &+ ({\omega}{r} - \frac{{\tilde{\alpha}}\mathcal{R}'}{\sqrt{\mathcal{R}^2 + {\mathcal{R}'}^2}})\vb*{e}_\theta,
\end{align}
where $\tilde{\chi} = \chi e^{-(t-T_c)/\tau}/[{\chi e^{-(t-T_c)/\tau} + 1-\chi}]$, $\tilde{H} = H(\mathcal{R}(\arg{(\vb*{r})}) - |\vb*{r}|)$, $\tilde{\alpha} = \alpha\Delta(\mathcal{R}(\arg{(\vb*{r})}) - |\vb*{r}|)$, and $\mathcal{R}' = \partial \mathcal{R}/\partial \theta$ are introduced for simplification.

\section{Programming Deformation through Modulation of Activity and Boundaries}

\begin{figure*}[t]
        \includegraphics[width=\linewidth]{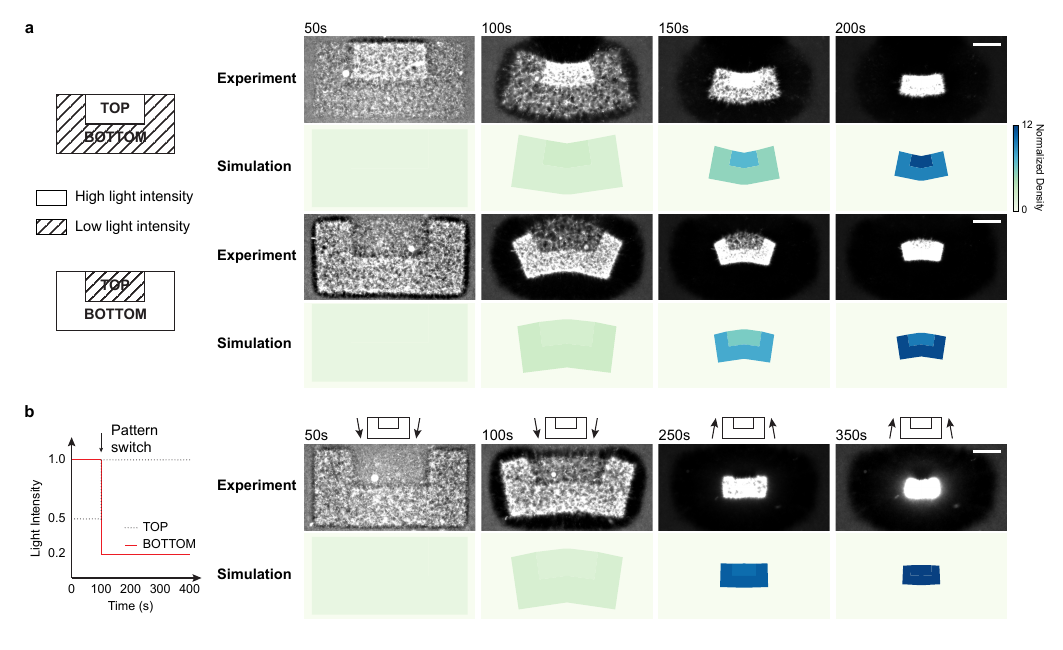}
        \caption{
         Contraction of the network can be programmed via modulating the pattern of illumination in space and time. (a) shows purely-spatial modulations of light, where the top segment of the rectangle is illuminated more/less strongly. Greater intensity of light activates larger number of motor proteins and thus generates larger active stresses, which leads to larger and faster contraction on the brighter side, and causes bending. Scale bar, 200$\mu$m. (b) The pattern of illumination varies in time to interpolate between the two static patterns of (a). Using this dynamic modulation we manage to change the bending direction as the network contracts. Scale bar, 200$\mu$m.
         } 
        \label{fig:fig-4}
\end{figure*}

Our theoretical model shows how the boundary programs network deformation by defining the domain of activity and providing boundary forces. However, programming deformation is difficult if the network only has a single predefined boundary. A simple strategy for programming the mechanical properties of MT networks is to divide network into multiple sub-networks by modulation of activity and boundaries. The strength of active stress $s$ depends on activity $a_{\mathcal{I}}$ and the boundary $\mathcal{R}$ by $s \propto a_{\mathcal{I}}\tilde{H}(\mathcal{R})$. Therefore, the boundaries of sub-networks can be created by spatial modulation of activity, i.e. $s \propto \sum_ia_{\mathcal{I}i}\tilde{H}(\mathcal{R}_i)$, where $a_{\mathcal{I}i}$ and $\mathcal{R}_i$ are the activity and boundary for a specific sub-network, corresponding to a region in the active network. By modulating light intensity in different regions, we redefine boundaries and allow the networks in different regions to contract according to their respective boundaries. It leads to novel mechanical behaviors that deviate from the contractions observed in networks at a uniform activity defined by a single boundary.

To demonstrate the programmable deformation through modulation of activity and boundaries, we designed a hinge light pattern for a rectangular MT network. In the pattern, the MT network is divided into two distinct regions and each region has a corresponding boundary. The differences in activity and boundary lead to relative differences in contractile velocity fields and network bending (Fig. \ref{fig:fig-4}a, top, Supplementary Video 9). In a complement hinge pattern, we induce bending along the opposite direction by switching the orientation of the joint (Fig. \ref{fig:fig-4}a, bottom, Supplementary Video 10). In addition to generating static deformations, spatial and temporal modulation of light patterns allow the generation of dynamic contraction and deformation through temporal modulation of relative activity. In particular, we temporally modulated the relative light intensity in the two regions of the hinge according to the following protocol (Fig. \ref{fig:fig-4}b, Supplementary Video 11). First we shine a light pattern that induces downward bending. The light pattern is subsequently swapped to the complementary pattern at around $t=100$s after the initial illumination. The differential intensities lead to reversal of the bending direction. The rates of the bending and reversal depend on the relative sizes of the two regions of illumination, relative light intensities, and the time at which swapping to complementary pattern takes place. Here we chose a relatively straightforward protocol with the same intensities and densities of MTs as chosen in the previously discussed cases. 

Broadly, these experiments show that both spatial-temporal modulation of light intensity allows us to redefine boundaries and induce programmed patterns of mechanical deformation into active MT networks. In this way, the natural contractile property of active MT networks can be simply modulated through relative differences in activity and boundary in distinct parts of an induced network. This controllability of MT networks allows us to program units of networks in which different possess engineered mechanical properties and can perform work in a programmed and predetermined manner through internal couplings. 

\section{Discussion}
\label{sec:discussion}

Active networks are ubiquitous in biology, and their non-equilibrium properties are poorly understood \cite{mogilner2018intracellular, alim2013random}. Our work reveals signature of activity in the mechanical properties at macroscopic scales. The active self-similar contraction is intrinsically related to the non-equilibrium nature of the system, which preserves a geometric memory, unlike in passive systems where equilibration increases entropy and erases the memory of the initial state. This memory preservation property makes the behavior of the system more controllable without the need to tuning the microscopic degrees of freedom. 

Previous works analyzed active contractions in networks of MT and actin in cell extracts, where the contracting network is embedded in a viscous solution, thus subjected to drag forces \cite{ross2019controlling, qu2021persistent}. Our optical control strategies allow us to isolate the networks from the surrounding solution while using light to modulate the boundaries and activity. Further, in conventional materials altering mechanical properties requires changing the microscopic structure of the material, for example, through doping \cite{li2020mechanical}. These changes are generically irreversible (plastic), and are hard to be modulated at the microscopic level \cite{meyers2008mechanical, keim2014mechanical}. In our systems, the degree of linking of the network and the active stresses can be tuned in space and time, enabling a separate strategy for the programming and control over material mechanics. Activity induced deformations provide a strategy for engineering novel behaviors at micron length scales.

Our work provides deeper insights into the important role of boundary geometry during network contraction, where the boundary not only directs the global contractile behavior but also generates local forces that cause the geometry distortion. However, some unanswered and exciting questions about the active matter system still exist. The network formation from the initially random orientation and the isolation process from the surrounding MTs prior to contraction require deeper investigation. Furthermore, the role of entropy reduction of active networks in the self-organizing process remains poorly understood. Our work offers insights and a useful method for dynamic programmable controlling active matters such as active fluids \cite{tayar2023controlling, shankar2022optimal, ghosh2024spatiotemporal, zhao2025dna}, active liquid crystals \cite{zhang2021spatiotemporal}, and soft robots \cite{blackert2025spatiotemporally, li2025photophosphorylation}. 

\bibliography{ref}

\end{document}